\documentclass{elsart3p}

\usepackage{graphicx}

\usepackage{amssymb}
\usepackage{textcomp}

\begin{document}

\begin{frontmatter}



\title{Simulation  of  a  hybrid  optical-radio-acoustic  neutrino  detector   at the South Pole}


\author[kansas]{D. Besson}, 
\author[desy]{R. Nahnhauer}, 
\author[berkeley]{P. B. Price}, 
\author[desy]{D. Tosi}, 
\ead{Delia.Tosi@desy.de} 
\author[berkeley]{J. Vandenbroucke},
\author[desy]{B.Voigt}

\address[kansas]{Dept. of Physics and Astronomy, University of Kansas, 
		Lawrence, KS 66045-2151, USA}
\address[desy]{DESY, Platanenallee 6, D-15738 Zeuthen, Germany}
\address[berkeley]{Dept. of Physics, University of California, Berkeley, CA 94720, USA}

\begin{abstract}
With construction halfway complete, IceCube is already the most sensitive neutrino
telescope ever built. A rearrangement of the final holes of IceCube with increased spacing has been discussed recently to optimize the high energy sensitivity of the detector. Extending this baseline
with radio and acoustic instrumentation in the same holes could further
improve the high energy response. The goal would be both to detect events and to act as a pathfinder for hybrid detection, towards a possible larger hybrid array. Simulation results for such an array are presented here.
\end{abstract}

\end{frontmatter}

\section{Introduction}
Detection of astrophysical neutrinos at energies above 10$^{16}$ eV would test cosmic ray
production models and fundamental particle physics at $\sim$100 TeV centre of mass energy. 

Techniques beyond optical measurements are required to detect 10 or more events per
year with a reasonable amount of instrumentation. Integrating both radio and acoustic
receivers with optical instrumentation could provide the necessary sensitivity as well as
allow for cross calibration and background rejection with hybrid events.
The full simulation of such a detector, combining optical, radio and acoustic detection methods, was done in \cite{VandenICRC} and revealed the possibility to detect $\sim$ 20 GZK neutrinos per year, 40\% of which were in the form of hybrid events detected by more than one technique. 

The optical Cherenkov detection technique is well established and fully exploited in IceCube, located in the deep ice at the South Pole and currently halfway constructed \cite{IceCube2008}. The radio and acoustic detection methods are currently in an intensive R\&D phase, with several experiments (RICE \cite{Besson08}, AURA \cite{AURA} and SPATS \cite{SPATS2007} \cite{SPATSArena08}) underway to measure \textit{in situ} the parameters necessary to establish the optimal geometry for a large detector. 

A first step towards the construction of a multi-km$^3$ hybrid detector could consist in a high energy hybrid extension around IceCube, combining radio and acoustic sensors with optical instrumentation. 
A simulation of a detector with such an extension has been performed to study what the improvement would be at highest energies.

\section{Simulation of a possible UHE extension to IceCube}
\subsection{Geometry}
\begin{figure}[h]
\centering
\includegraphics[width=7.5cm,height=5cm]{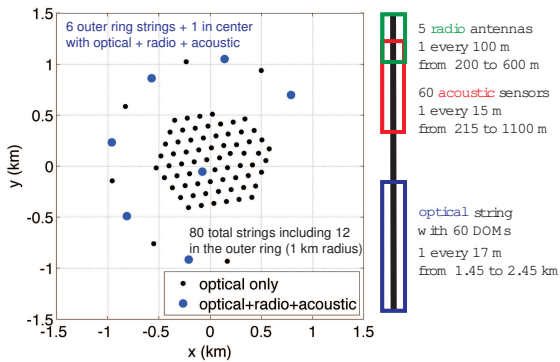}
\caption{Geometry considered for the UHE extension to the IceCube detector. Simulated instrumentation, as a function of the depth along a single string, is schematically indicated on the right.}
\label{fig:1}
\end{figure}

IceCube will be made of 80 strings arrayed hexagonally with 125 m horizontal spacing. Each string is equipped with 60 DOMs (Digital Optical Modules), with 17 m vertical spacing at depths between 1.45 km and 2.45 km. In the geometry simulated, 12 of the strings are rearranged at 1 km distance from the centre, taking into account the presence of the skiway on one side of the detector, which does not allow for drilling \cite{KarleNeutrino08}. Half of these strings, plus one at the centre of the array, are equipped additionally with other technologies: 5 radio antennas are placed every 100 m from 200 m to 600 m depth, and 60 acoustic sensors are located every 15 m from 215 m to 1100 m depth. An illustration of the geometry is shown in \figurename{ \ref{fig:1}}.

\subsection{Simulation Methods}
The optical simulation is run with standard IceCube software \cite{IceCubeSim}, which takes into account particle propagation for muons and electrons and the Landau-Pomeranchuk-Migdal (LPM) effect. Light is propagated using an ice model with dust layers \cite{IceModel}. The DOMs are simulated as realistically as possible. Both a global and a local trigger condition are set, requiring a minimum of 8 modules in the whole array being hit in 5 ${\mu}$s and at least two adjacent DOMs being hit in 1 ${\mu}$s.

For the radio simulation, the signal from a shower is modelled as in \cite{Alvarez2000}, including the LPM effect as prescribed in \cite{Alvarez1997}, with the assumption of a maximum field attenuation length of 1.2 km. Every radio receiver is a single half-wave dipole antenna with an effective height of 0.27 m at the frequency of 200 MHz. A Gaussian bandwidth with a $\sigma\sim$60 MHz and a sharp high pass filter at 110 MHz are included in the receiver model. The trigger condition applied requires a signal above 4.5 $\sigma_{kT}$ in 4 antennas, where $\sigma_{kT}$ is the rms noise (including thermal and 250 K system temperature). 

The acoustic simulation uses a modified version of the Nishimura-Kamata-Greisen (NKG) parameterization for hadronic showers, which includes lengthening by the LPM effect \cite{NKGmod} \cite{Alvarez1998}.
Electromagnetic showers generated from primary or secondary electron neutrinos are expected to be strongly elongated by the LPM effect, causing a weaker acoustic signal; therefore they are ignored. Given the energy deposited at each interaction, the bipolar acoustic waveform is simulated in a way similar to that used for SAUND \cite{Vandenbroucke2005}. The signal is calculated taking into account a depth-dependent attenuation model \cite{Price2006}, which predicts an attenuation length between 7.5 km and 4.5 km at the instrumented depths. A uniform frequency and angular response of the sensor is assumed. The trigger condition requires a signal above 9 mPa recorded by at least 3 sensors. The latest measurements made \textit{in situ} with SPATS compared to calibration data in the laboratory show that this assumption is reasonable \cite{KargArena08}.

\section{Simulation results}
\subsection{Neutrino Set}
A set of 10$^6$ neutrinos (including neutrinos and antineutrinos of all 3 flavours ${\nu_{\e}}$, ${\nu_{\mu}}$, ${\nu_{\tau}}$), following a \textit{E${^{-1}}$} spectrum from 10${^{16.5}}$ eV to 10${^{19.5}}$ eV, has been produced using the neutrino generator ANIS \cite{ANIS}. This includes propagation through air, rock and ice, with CTEQ5 cross sections to calculate neutral and charged current interaction rates. Neutrinos were generated up to a zenith angle of 120$^o$ since at these energies upgoing neutrinos are absorbed by the Earth.

The generation volume was a cylinder of 10 km radius, 20 km length, resulting in a volume of $\sim$6280 km$^3$. This volume is rotated in the direction of each neutrino. All the neutrinos reaching the generation volume are forced to interact and a weight is assigned to take into account the interaction probability. Each interaction can produce a signal detectable by any of the techniques which are simulated independently. The identification of the event is unique so it is possible to identify events detected by any combination of methods. 

\begin{figure}[ht]
\centering
\includegraphics[width=7.5cm,height=6.5cm]{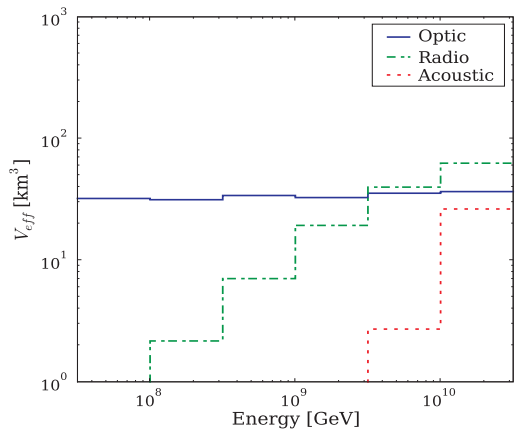}
\caption{Effective volume for the optical, radio and acoustic detection method in the energy range of the simulated neutrino sample.}
\label{fig:effVol}
\end{figure}

The number of events detected by each single method and by each combination of methods has been used to calculate the effective volume as a function of energy. The results are shown in \figurename{ \ref{fig:effVol}}. The flatness of the optical curve stems from the contribution of muons to the signal events. At the energy studied here, due to the limited size of the simulated interaction volume, the likelihood of a muon to reach the detector is independent of the energy. Acoustic and radio detection methods are instead sensitive to signals produced by hadronic showers with a particular threshold and a strength increasing with the energy. 

\begin{figure}[ht]
\centering
\includegraphics[width=7.5cm,height=6.5cm]{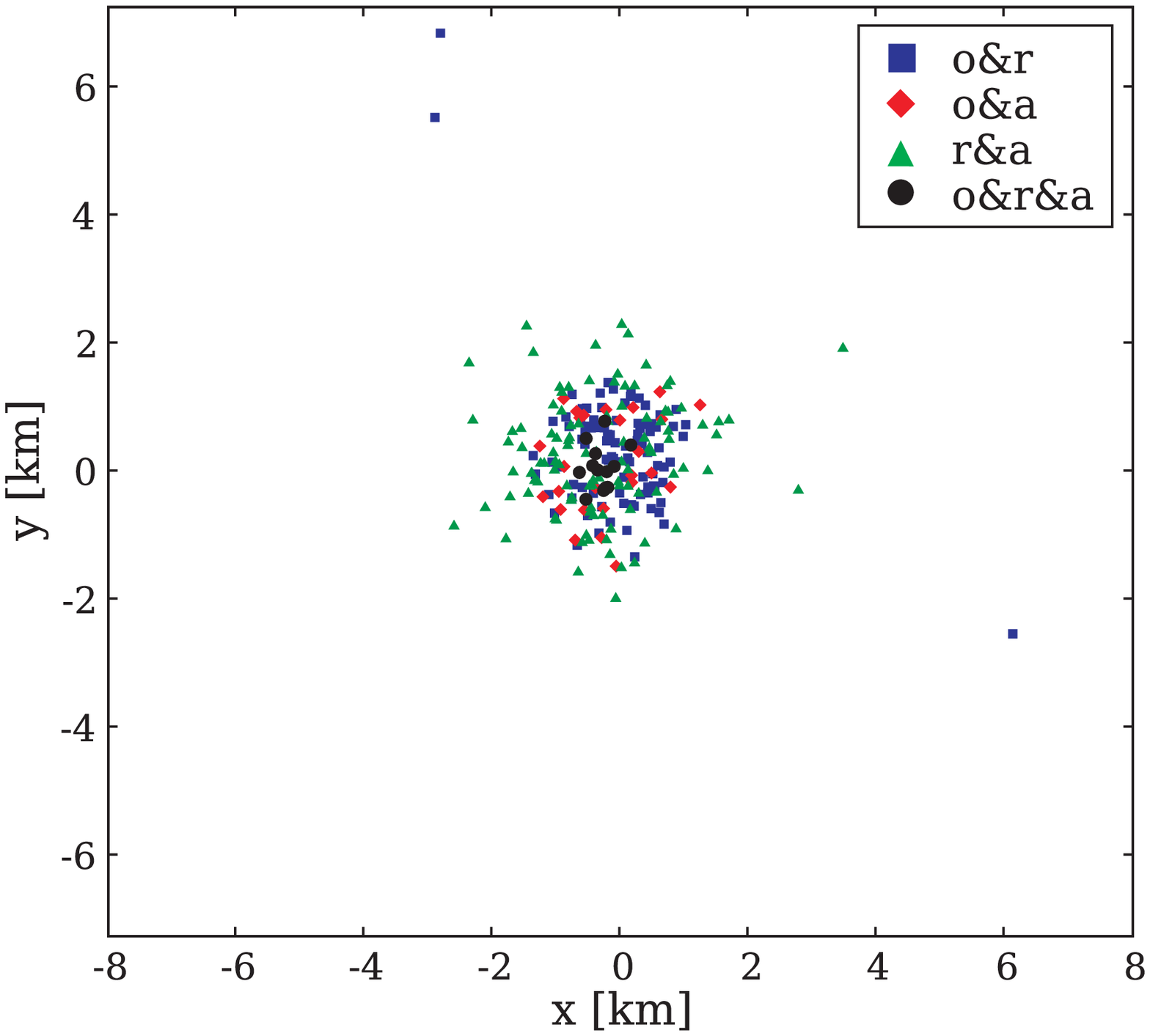}
\caption{Distribution of the hybrid events in the \textit{x}, \textit{y} coordinates.}
\label{fig:xy}
\end{figure}

\begin{figure}[ht]
\centering
\includegraphics[width=7.5cm,height=6.5cm]{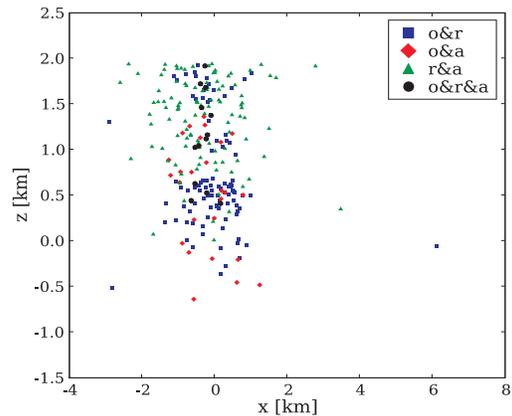}
\caption{Distribution of the hybrid events in the \textit{x}, \textit{z} coordinates.}
\label{fig:xz}
\end{figure}

The distribution of hybrid events detected by more than one method is shown in \figurename{ \ref{fig:xy}} and in \figurename{ \ref{fig:xz}} using the IceCube coordinate system, whose origin is located in the centre of the detector, 1948.07 m below the surface. The horizontal coordinates are (x, y) and z is positive upward.


The resulting effective volumes have been folded with the Engel-Seckel-Stanev GZK flux model \cite{ESS}, assuming a source evolution parameter $\Omega_\Lambda$=0.7. Interactions by $\nu_\tau$ and $\overline{\nu_\tau}$ are introduced in this model assuming a resulting final mixing 1:1:1 due to flavour oscillation. 

The number of GZK neutrinos detected per year by each single method and each combination of methods is shown in \tablename{ \ref{tab:1}}. The total number has been calculated considering all the events detected by any method. 

\begin{figure}[ht]
\centering
\includegraphics[width=7.5cm,height=6.5cm]{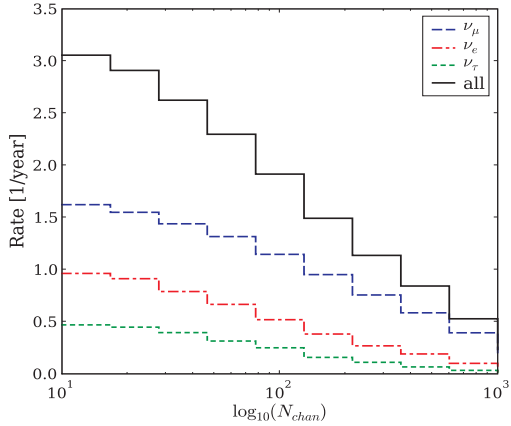}
\caption{Rate of events in the optical channel vs. the number of channels hit chosen as trigger threshold.}
\label{fig:optic_rates_vs_nchan}
\end{figure}

Reasonable background rejection may require a cut on the optical channel multiplicity. With $\mathit{NCh}$ $>$ 100 the number of optically observable events would be reduced by less than a factor of $\sim$2, as shown in \figurename{ \ref{fig:optic_rates_vs_nchan}}.

\begin{table}[htbp]
\begin{center} 
\caption{GZK rates per year from the simulated geometry. ``Optical'' refers to IceCube plus the optical channel of the extension.}
\label{tab:1}
\begin{tabular}{c|c}
\textbf{Detection option} & \textbf{GZK events per year} \\ \hline \hline
 IceCube & 1.78 \\ \hline
 Optical & 3.09 \\ \hline
 Radio	& 1.31 \\ \hline
 Acoustic & 0.16 \\ \hline
 Optical + Radio & 0.15 \\ \hline
 Optical + Acoustic & 0.03 \\ \hline
 Radio + Acoustic & 0.08 \\ \hline
 Optical + Radio + Acoustic & 0.01 \\ \hline
\textbf{Total} & 4.32 \\ 
\end{tabular}
\end{center}
\end{table}

\section{Conclusion}
The simulated extension increases the number of GZK neutrino events seen by IceCube by a factor of $\sim$2. 
The detection of a few additional events by the radio method would strengthen the observation. The detector extension in the present form would be too small to profit from acoustic measurements. However in the present effort neither the string nor the sensor geometry nor the trigger conditions have been optimized for the radio and acoustic detector components. 

An optimized medium size extension of IceCube could nevertheless build the bridge to a large volume, high sensitivity GZK neutrino detector. It would also be a valuable prototype to test and develop the most suitable technology and infrastructure. 

\section{Acknowledgment}
The support of this work by the IceCube Collaboration and NSF is kindly acknowledged. 



\end{document}